\documentclass[a4paper,10pt,twocolumn]{article}

\usepackage[utf8]{inputenc}
\usepackage[english]{babel}
\usepackage{graphicx}
\usepackage{dblfloatfix}
\usepackage{placeins}
\usepackage{etoolbox}

\pretocmd{\subsection}{\FloatBarrier}{}{}

\usepackage{caption}
\usepackage{geometry}
\usepackage{float}

\usepackage{csquotes}
\usepackage[
  backend=biber,
  style=numeric-comp,
  sorting=none
]{biblatex}

\usepackage{hyperref}

\begin{document}

\begin{refsection}

\onecolumn

\begin{center}
    \centering
    \Large\textbf{Multi-wavelength UV Upconversion in Lanthanides assisted by Photonic Crystals}\\
    \vspace{0.5cm}
    \normalsize Damien Rinnert\textsuperscript{1,2,*}, Emmanuel Drouard\textsuperscript{1}, Antonio Pereira\textsuperscript{2}, Céline Chevalier\textsuperscript{1}, Aziz Benamrouche\textsuperscript{1}, Benjamin Fornacciari\textsuperscript{1}, Hai Son Nguyen\textsuperscript{1}, Gilles Ledoux\textsuperscript{2,*}, and Christian Seassal\textsuperscript{1,*}\\
    \vspace{0.3cm}
    \small \textsuperscript{1} CNRS, INSA Lyon, Ecole Centrale Lyon, Université Claude Bernard Lyon 1, CPE Lyon, INL, UMR 5270, F-69622, Villeurbanne, France\\
    \vspace{0.15cm}
    \small \textsuperscript{2} Université Claude Bernard Lyon 1, CNRS, Institut Lumière Matière, UMR5306, F-69100, Villeurbanne, France\\
    \vspace{0.15cm}
    \small \textsuperscript{*} \textit{damien.rinnert@ec-lyon.fr; gilles.ledoux@univ-lyon1.fr; christian.seassal@ec-lyon.fr}
\end{center}

\vspace{0.5cm}

\noindent\textbf{Abstract:} Upconversion luminescence consists of the absorption of low-energies photons followed by the emission of a higher energy photon. The process has mainly been studied in lanthanides to upconvert monochromatic near-infrared excitation to near-infrared or visible light, and has been exploited only to a limited extent to upconvert broad excitations to ultra-violet. In addition, upconverting near-infrared and visible light to ultra-violet is crucial for applications such as solar-to-fuel conversion or environmental remediation. However, upconversion luminescence is limited by the low absorption cross-sections of lanthanides. In this work, we engineered Bloch modes in a photonic crystal to assist a multi-wavelength upconversion mechanism and demonstrated a 28-fold enhancement of ultra-violet upconversion luminescence of Yb\textsuperscript{3+}-Tm\textsuperscript{3+} doped thin films. Materials were selected and optimized to design nanostructures without parasitic absorption losses. The geometric parameters of the photonic crystals were scanned to match a slow-light resonance with an excited-state transition of Tm\textsuperscript{3+} and thus enhance incident visible light absorption. Ultra-violet light extraction was also enhanced by photonic crystal Bloch modes. Each of these two contributions were quantified and the measured photonic band structures were well reproduced by electromagnetic simulations.\\


\noindent\textbf{Abbreviations:} upconversion luminescence (UCL), ultra-violet (UV), near-infrared (NIR), photonic crystal (PhC), energy transfer (ET), excited-state absorption (ESA), local photonic density of states (LDOS), scanning electron microscope (SEM), pulsed laser deposition (PLD), plasma enhanced chemical vapor deposition (PECVD), rigorous coupled-wave analysis (RCWA), inductively coupled plasma - reactive ion etching (ICP-RIE), numerical aperture (NA)

\twocolumn

\section*{Introduction}
%%%%    From UV upconversion requirement to the inherent low absorption cross-section of lanthanides, passing by UC mechanisms in Yb3+ - Tm3+ co-doped system

Upconversion luminescence (UCL) is a fluorescence process consisting of the sequential absorption of several low-energy photons followed by the emission of higher-energy photons. Although it has been extensively studied in recent years for its ability to convert near-infrared (NIR) to higher energies \cite{RSC_Wang_2009, NatNano_Zhou_2015, ACS_Zhou_2015}, the process has only been exploited to a limited extent to upconvert broad excitations such as solar radiation to a specific wavelength \cite{ACS_Purohit_2019}. In addition, the upconversion from NIR and visible light to UV range is driven by solar energy and environmental applications such as the generation of solar fuels, pollution-reducing and self-cleaning surfaces \cite{NatEnergy_Wang_2022, RSC_Xue_2024, Makiyama_JNR_2025, RSC_Magazov_2026}.

\begin{figure}[!h]
    \centering
    \includegraphics[]{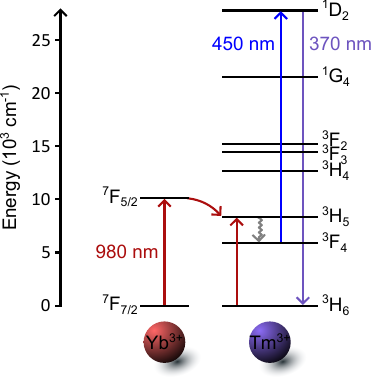}
    \caption{Multi-wavelength UV UCL mechanism in Yb\textsuperscript{3+}-Tm\textsuperscript{3+}. It involves 1 NIR photon absorption by Yb\textsuperscript{3+} (\textsuperscript{7}F\textsubscript{7/2} $\rightarrow$ \textsuperscript{7}F\textsubscript{5/2}), whose energy is transferred to Tm\textsuperscript{3+} (\textsuperscript{3}F\textsubscript{4}), followed by 1 visible excited state absorption in Tm\textsuperscript{3+} (\textsuperscript{3}F\textsubscript{4} $\rightarrow$ \textsuperscript{1}D\textsubscript{2}). UV emission is obtained from spontaneous emission to the ground-state (\textsuperscript{1}D\textsubscript{2} $\rightarrow$ \textsuperscript{3}H\textsubscript{6}).}
    \label{fig:UC_process}
\end{figure}

Different systems have been considered in order to provide UV upconversion. Triplet-triplet annihilation has been intensively investigated in molecules during the last few years as a potential candidate \cite{ACS_Olesund_2022, RSC_Uji_2022}. Although encouraging efficiencies have been obtained, it currently remains limited to small anti-Stokes shifts of up to 1eV due to the significant energy losses that occur during the upconversion process \cite{ACS_Wang_2025}. Consequently, it prevents these systems from harvesting NIR photons. In contrast, lanthanides enable substantial anti-Stokes shifts with considerable photostability and long luminescence lifetimes \cite{CR1_Auzel_1966, CR2_Auzel_1966, ACS_Auzel_2004, ACS_Villanueva_2015, ACS_Purohit_2019, ACS_Li_2025}. In particular, the co-doped system Yb\textsuperscript{3+}-Tm\textsuperscript{3+} can upconvert NIR and visible photons into UV photons. So far, most of the upconversion mechanisms studied are based on energy transfer (ET) from Yb\textsuperscript{3+} (sensitizer) to Tm\textsuperscript{3+} (activator) under NIR ($\sim$980 nm) excitation. Although spontaneous emission Tm\textsuperscript{3+} (\textsuperscript{1}D\textsubscript{2} $\rightarrow$ \textsuperscript{3}H\textsubscript{6}) of one UV photon ($\sim$370 nm) can be obtained from three to four successive ETs (Supplemental Information \ref{app:UC}) \cite{NatMat_Wang_2011, RSC_Mishra_2012, ACS_Purohit_2019}, these mechanisms have mainly been addressed to obtain NIR to NIR and NIR to visible UCL \cite{DeGruyter_Das_2020}. In 2019, B. Purohit et al. showed that these ET based mechanisms do not generate detectable UV emission under sun-like NIR irradiance ($\sim$15 mW/cm\textsuperscript{2}). The authors investigated a much less studied mechanism based on multi-wavelength excitation: NIR ($\sim$980 nm) and visible ($\sim$450 nm). The latter is of great interest as it generated detectable UV emission under solar-like irradiances: NIR ($\sim$15 mW/cm\textsuperscript{2}) and visible ($\sim$3 mW/cm\textsuperscript{2}) \cite{ACS_Purohit_2019}. This mechanism is presented in Figure \ref{fig:UC_process}. First, Yb\textsuperscript{3+} is excited by the absorption of one NIR photon. Then Förster energy transfer from Yb\textsuperscript{3+} to Tm\textsuperscript{3+} followed by non-radiative relaxation populates Tm\textsuperscript{3+}(\textsuperscript{3}F\textsubscript{4}) excited-state, whose lifetime is higher than 10 ms \cite{ACS_Villanueva_2015}. Next, excited-state absorption (ESA) of one visible photon excites Tm\textsuperscript{3+}(\textsuperscript{1}D\textsubscript{2}), which is the strongest electric-dipole interaction from the excited state \textsuperscript{3}F\textsubscript{4} \cite{ANL_Carnall_1978}. UV emission is finally obtained by spontaneous emission to the ground-state of Tm\textsuperscript{3+} (\textsuperscript{3}H\textsubscript{6}). The presented mechanisms characterize different UCL regimes that depend on the local excitation irradiance. They are ruled in weak excitation regimes by Equation \ref{eq:I_UCL}, where non-linear orders $n_{NIR}$ and $n_{vis}$ are, respectively, the number of NIR and visible photons involved in the upconversion mechanism \cite{DeGruyter_Das_2020}.
\begin{equation}
    I_{UCL} \propto \left(I_{NIR}^{loc}\right)^{n_{NIR}} \times \left(I_{vis}^{loc}\right)^{n_{vis}}
    \label{eq:I_UCL}
\end{equation}
However, the UCL efficiency of lanthanides is limited by the forbidden nature of the 4f-4f transitions involved, and therefore low absorption cross-sections ($\sigma_{abs}\sim 10^{-20}$ cm\textsuperscript{2}) \cite{JoL_Hehlen_2013, JoL_Moncorge_2022}, cross-relaxations and surface quenching.

%%%%    State-of-the-art: Enhancing Ln3+ UCL performances, material and photonic approaches

Multiple approaches have been studied to improve the UCL efficiency of lanthanide. On the one hand, materials have been optimized to promote Förster energy transfers and reduce surface quenching with suitable host matrices, core-shell nanostructures, and optimizations of doping rates \cite{RSC_Wang_2009, NatMat_Wang_2011, NatNano_Zhou_2015, ACS_Zhou_2015}. On the other hand, since the absorption cross-section scales with the square of the local electromagnetic field ($\sigma_{abs}\propto |E|^2$), photonic approaches have been proposed to engineer the local photonic density of states (LDOS) and thus increase the light-matter interaction in upconverting nanoparticles: lensing \cite{NatCom_Liang_2019}, plasmon resonances \cite{ACS_Li_2024, DeGruyter_Das_2020}, opal photonic crystal \cite{RSC_Yin_2013, JMatSci_Shi_2019}, and photonic crystal slabs (PhC slabs) \cite{RSC_Wang_2016, ACS_Lin_2015, ACS_Vu_2018, ACS_Mao_2019, Nanoscale_Gong_2019, ACS_Würth_2020, AOM_Ahiboz_2021, Crystals_Vu_2021, ACS_Gao_2023, Nat_Schiattarella_2024, ACS_Tseng_2024}. The latter benefit from limited parasitic absorption losses and quenching. Additionally, they offer multiple degrees of freedom, which enable optical resonances control provided that their geometry is carefully adjusted. In particular, PhC slabs exhibit Bloch modes, which are non-local optical modes that can be efficiently addressed by out-of-plane incident light. A Bloch mode whose dispersion verifies $d\omega/dk_{\parallel}=0$ is called a slow-light resonance, where $\omega$ is the energy of the mode and $k_{\parallel}$ the in-plane component of the incident light wave vector.  PhC slabs patterned as a square lattice provide such slow-light resonances at normal incidence, which corresponds to the so called $\Gamma$-point in the center of the first Brillouin zone \cite{ACS_Mao_2019, ACS_Würth_2020, ACS_Gao_2023, Nat_Schiattarella_2024, ACS_Yuan_2024}. By trapping excitation light, they can increase the local irradiance by one order of magnitude \cite{ACS_Lin_2015, ACS_Vu_2018, Crystals_Vu_2021}. As a result, these slow-light resonances at the excitation wavelength can enhance the UCL by a factor of 10\textsuperscript{n\textsubscript{ph}}, where n\textsubscript{ph} is the number of photons involved in the upconversion process at the excitation wavelength considered (Equation \ref{eq:I_UCL}). Moreover, a Bloch mode at the emission wavelength can increase the extraction of light for a given solid angle by a factor of two to four \cite{ACS_Lin_2015, ACS_Vu_2018, Crystals_Vu_2021}. Several authors have used this approach to assist the UCL from NIR to visible through ET based mechanisms in the Yb\textsuperscript{3+}-Er\textsuperscript{3+} system \cite{RSC_Wang_2016, ACS_Mao_2019, ACS_Würth_2020, AOM_Ahiboz_2021, ACS_Gao_2023}, the Yb\textsuperscript{3+}-Tm\textsuperscript{3+} system \cite{ACS_Lin_2015, ACS_Vu_2018}, or both \cite{Crystals_Vu_2021, Nat_Schiattarella_2024, ACS_Tseng_2024}. Nevertheless, to the best of our knowledge, the potential of LDOS engineering has not been explored for multi-wavelength regimes in lanthanides, and in particular for UV UCL from NIR and visible.

%%%%    Our approach, main results and methods

In contrast of previous studies on LDOS engineering for UCL enhancement in lanthanide-doped nanoparticles, we propose a PhC structure designed for NIR, visible and UV ranges to address the multi-wavelength regime by combining a lanthanide-doped thin film with a patterned transparent thin film. Deposition techniques enable a good control over the thicknesses of thin films, and thus over Bloch modes. We achieve a 28-fold enhancement of the UV UCL by increasing both the visible absorption (\textsuperscript{3}F\textsubscript{4} $\rightarrow$ \textsuperscript{1}D\textsubscript{2}) and the UV extraction (\textsuperscript{1}D\textsubscript{2} $\rightarrow$ \textsuperscript{3}H\textsubscript{6}). To achieve this, slow-light resonance is tuned with the (\textsuperscript{3}F\textsubscript{4} $\rightarrow$ \textsuperscript{1}D\textsubscript{2}) transition, while the PhC structure further controls UV extraction with Bloch modes. We fabricated PhC structures with lattice parameters in the 240 - 280 nm range by a top-down approach. Their dispersion characteristics were measured and showed good agreement with our simulations. The multi-wavelength upconversion mechanism has been confirmed with measurements of UCL power laws. Finally, contributions of both visible absorption and UV extraction to the global enhancement of UV UCL have been experimentally characterized.

\section*{Results \& Discussions}
\begin{figure}
    \centering
    \includegraphics{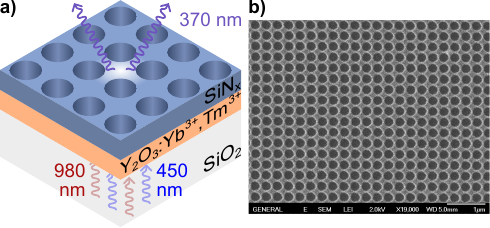}
    \caption{Presentation of the PhC structure. (a) 3D scheme of the PhC design and operational principle: NIR and visible light absorption close to normal incidence and UV emission in a controlled direction. (b) SEM top view image of a fabricated PhC with the following parameters: period p = 267.3 $\pm$ 0.4 nm; air filling factor ff = 47.9\% $\pm$ 0.2\%; etching depth H = 138 $\pm$ 1 nm.}
    \label{fig:design_illustration}
\end{figure}

%%%%    Design principle & fabrication

The design and a scanning electron microscope (SEM) top-view image of the PhC structure are presented in Figures \ref{fig:design_illustration}a \& \ref{fig:design_illustration}b. Doping a matrix of yttrium oxide (Y\textsubscript{2}O\textsubscript{3}) and adjusting the stoichiometry of the silicon nitride layer leads to a PhC structure that is transparent from the UV to the NIR ranges, except for the absorption of the lanthanide co-doped system. First, a 95 nm thick thin film of doped yttrium oxide, Y\textsubscript{2}O\textsubscript{3}:Yb\textsuperscript{3+}(7.5\%),Tm\textsuperscript{3+}(0.5\%), was deposited on top of a fused silica wafer by pulsed laser deposition (PLD). A home-made PLD target with doping rates that have been optimized to maximize UV emission was used. As expected for the Y\textsubscript{2}O\textsubscript{3} matrix, the crystalline phase of the PLD target was measured to be cubic with X-ray diffraction method \cite{MatSciFor_Baldinozzi_1998} (Supplemental Information S1-C.1.1). The spectral shape of the ESA transition \textsuperscript{3}F\textsubscript{4} $\rightarrow$ \textsuperscript{1}D\textsubscript{2} was experimentally obtained by measuring the excitation spectra of the PLD target while pumping the \textsuperscript{3}F\textsubscript{4} excited-state (Supplemental Information S1-C.1.1). Then a 144 nm thick SiN\textsubscript{x} thin film was deposited by plasma enhanced chemical vapor deposition (PECVD). Its stoichiometry has been optimized to meet the transparency requirements while maintaining a refractive index as high as possible. Its value at 450 nm was measured by ellipsometry to be n\textsubscript{SiN\textsubscript{x}} = 1.87, which is close to that of Y\textsubscript{2}O\textsubscript{3} (n\textsubscript{Y\textsubscript{2}O\textsubscript{3}} = 1.83). Using rigorous coupled-wave analysis simulations (RCWA) \cite{CPC_Liu_2012}, the depth (H) and radius (r) of the PhC holes, and the square lattice period (p) were scanned in order to tune the excited-state absorption transition \textsuperscript{3}F\textsubscript{4} $\rightarrow$ \textsuperscript{1}D\textsubscript{2} with a slow-light resonance in the $\Gamma$-point (Supplemental Information S1-B.1). The simulated dispersion of the resulting PhC structure is devoid of Bloch modes at lower energies in the $\Gamma$-point, and therefore in the NIR range. In addition, it exhibits Bloch modes in the UV emission range at higher incidence angles that are expected to assist light extraction. Next, 22 areas of 60 microns x 60 microns size of hole square lattices were defined by electron-beam lithography (EBL). The lattice parameter was scanned close to the value identified by our simulations, in the range of 240 - 280 nm, while the air filling factor was set to $ff=\pi r^2/p^2=50\%$. Finally, patterns were transferred throughout the entire SiN\textsubscript{x} layer using inductively coupled plasma - reactive ion etching (ICP-RIE).

%%%%    Optical characterization - UCL

\begin{figure*}[t]
    \centering
    \includegraphics{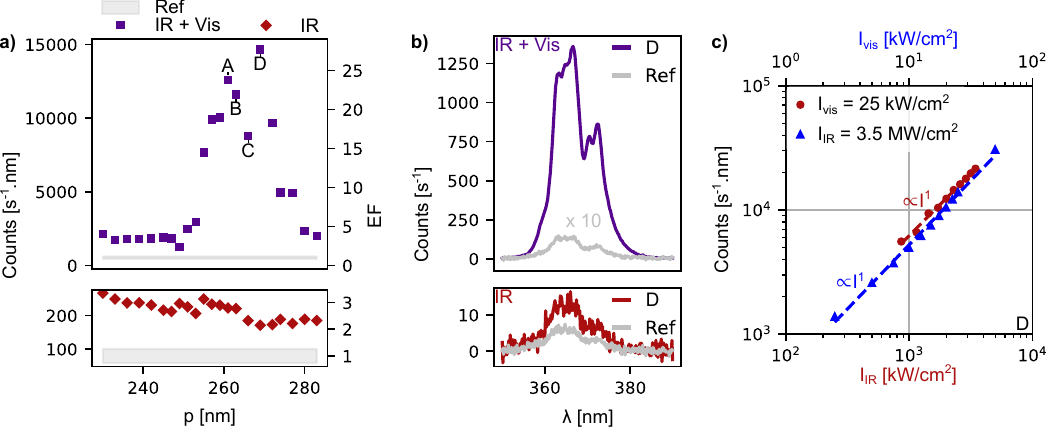}
    \caption{UV UCL experimental characterisation. (a-b) Measurements were done under NIR excitation ($\lambda$\textsubscript{NIR} =  995 nm | NA\textsubscript{NIR} = 0.40), with (top purple plots) or without (bottom red plots) the visible excitation ($\lambda$\textsubscript{vis} = 447 nm | NA\textsubscript{vis} = 0.04). UV emission is collected with a microscope objective of numerical aperture NA\textsubscript{UV} = 0.75. Considered irradiance are I\textsubscript{NIR} = 966.5 kW/cm\textsuperscript{2} and I\textsubscript{vis} = 18.2 kW/cm\textsuperscript{2}. (a) UV emission of 22 PhCs integrated with the composite trapezoidal rule depending on the PhC lattice parameter, and compared to references. Gray bands are reference values at 95\% confidence interval considering a normal distribution in the population of 5 measured unpatterned areas. PhC structures A, B, C and D are identified. (b) Emission spectra of the PhC D is compared to the averaged UCL signal of the references (gray curves). UCL data considered for the reference are obtained by averaging responses of 5 unpatterned areas. UCL data of the reference under IR and visible excitations is multiplied by a factor of 10 for the sake of visibility. (c) Evolution of the integrated UV emission of the structure D depending on the excitation conditions: NIR irradiance varies while visible one is fixed at 25 kW/cm\textsuperscript{2} (red dots), or visible irradiance varies while NIR one is fixed at 3.5 MW/cm\textsuperscript{2} (blue triangles). Dashed lines results from linear regressions.}
    \label{fig:raw_UC}
\end{figure*}

The UCL of all PhCs were measured and compared to that of the unpatterned layer stack, here considered as a reference. The samples were excited from the backside using collimated continuous-wave laser beams at 995 nm and 447 nm focused through a microscope objective of numerical aperture NA = 0.4. The size of the visible beam was reduced using a telescope to limit its effective numerical aperture of excitation to NA\textsubscript{vis} = 0.04. It enabled the setup to selectively address slow-light resonances in the $\Gamma$-point. UV emission was collected from the top side using a microscope objective with a numerical aperture NA\textsubscript{UV} = 0.75. We introduce the enhancement factor (EF) as the ratio of the integrated number of UV photons collected in the 350 - 390 nm spectral range from a PhC over the number of integrated UV photons collected from the reference in the same spectral range and under the same excitation conditions. Figure \ref{fig:raw_UC}a summarizes PhCs and reference UCL performances. Under NIR excitation, the PhC EF ranges from 2 to 4, without any particular tendency. In contrast, when excited by both NIR and visible light, UV emission peaks for lattice parameters of 259 nm (referred as A) and 267 nm (referred as D), and EF ranges from 2 to 28. The UV emission spectra of structure D, which provides the highest UCL, are shown and compared to those of the reference in Figure \ref{fig:raw_UC}b. Repeatability measurements were carried out on this PhC. An EF of 2.71 $\pm$ 0.5 is measured under NIR excitation and it reaches 28.1 $\pm$ 3.0 under both NIR and visible excitations.

%%%%    Optical characterization - Power law

Furthermore, power laws of the UV UCL of PhC D were measured to determine the upconverting mechanism involved (Figure \ref{fig:raw_UC}c). Non-linear orders are determined to be $n_{NIR} = n_{vis} = 1$ by fitting the experimental data with Equation \ref{eq:I_UCL}. Determination coefficients r\textsuperscript{2} obtained are higher than 0.99 (Supplemental Information S1-D.2). Thus, it confirms that the UV UCL of these structures occurs according to the multi-wavelength upconversion mechanism presented in Figure \ref{fig:UC_process}.

%%%%    Optical characterization - band structure

\begin{figure*}[t]
    \centering
    \includegraphics{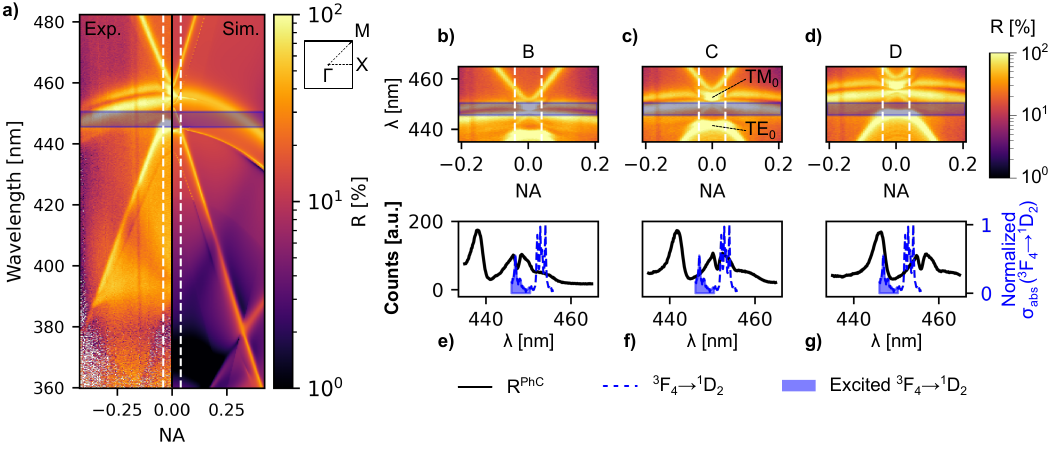}
    \caption{Band structure measurements and analysis along the high symmetry direction $\Gamma$X. Reflectivity measurements of PhC slabs have been conducted using a halogen-deuterium white light source and a microscope objective of numerical aperture NA\textsubscript{BS} = 0.28. White doted lines delimit incidence angles covered by NA\textsubscript{vis} from the UCL measurement setup. Blue doted lines correspond to the normalized absorption cross-section of the \textsuperscript{3}F\textsubscript{4} $\rightarrow$ \textsuperscript{1}D\textsubscript{2} transition. Blue areas correspond to the spectral range of the transition excited by the laser used in the UCL setup. (a) Measured band structure of the PhC D (left-side) compared to the RCWA simulation (right-side). (b), (c), and (d) are experimental band structure truncations centered on the $\Gamma$-point at \textsuperscript{3}F\textsubscript{4} $\rightarrow$ \textsuperscript{1}D\textsubscript{2} transition of B, C and D PhC structures respectively. TE\textsubscript{0}-like and TM\textsubscript{0}-like slow-light resonances are identified on the band structure (c). (e), (f) and (g) are reflectance spectra averaged over NA\textsubscript{vis}, denoted R\textsuperscript{PhC}, of B, C and D PhC structures respectively, overlapped by the normalized absorption cross-section $\sigma_{Tm^{3+}}$.}
    \label{fig:BFP}
\end{figure*}

\begin{figure}
    \centering
    \includegraphics{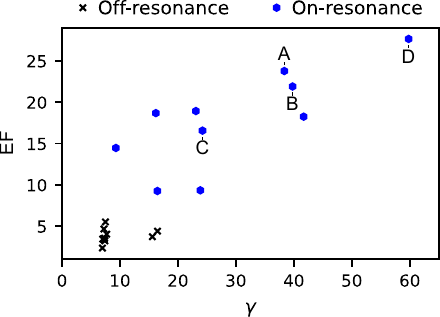}
    \caption{UV emission enhancement factor under NIR and visible excitation depending on the overlap factor $\gamma$.}
    \label{fig:FOM}
\end{figure}

To experimentally identify Bloch modes of all PhC, reflectivity measurements were performed along the high symmetry direction $\Gamma$X with a Fourier imaging setup. The measured band structure of PhC D is consistent with our simulation, as shown in Figure \ref{fig:BFP}a. Two slow-light resonances are identified at the $\Gamma$-point, close to the \textsuperscript{3}F\textsubscript{4} $\rightarrow$ \textsuperscript{1}D\textsubscript{2} transition wavelength. The low energy one is a TM\textsubscript{0}-like slow-light resonance while the second is a TE\textsubscript{0}-like slow-light resonance (Figure \ref{fig:BFP}c). As expected, Bloch modes are observed in the UV emission range at non-zero angles. Scanning the lattice parameter enables fine tuning of the band structure. Indeed, the latter is redshifted in the spectral range of interest by 1.39 $\pm$ 0.01 nm as the lattice parameter increases by 1 nm (Supplemental Information S1-D.5). No matter the lattice parameter in the covered range, band structures are devoid of Bloch modes in the NIR range at the $\Gamma$-point and show Bloch mode in the UV emission range. Their emission directions vary, but remain contained in the numerical aperture of collection NA\textsubscript{UV}. Hence, UCL enhancements measured under NIR excitation (Figures \ref{fig:raw_UC}a - \ref{fig:raw_UC}b) are the result of UV extraction assisted by patterning, especially through these Bloch modes. In contrast, the presence of slow-light resonances in the visible ESA range at the $\Gamma$-point is very sensitive to variations in the lattice parameter. Indeed, visible excitation is limited to the small numerical aperture NA\textsubscript{vis}. As a result, only PhC slabs with lattice parameter in the 253 - 275 nm range are resonant under these excitation conditions. To emphasize the impact of this sensitivity on the UCL, Figures \ref{fig:BFP}b - \ref{fig:BFP}g focus on PhC structures B, C and D. Figures \ref{fig:BFP}b - \ref{fig:BFP}d present their band structures. Taking into account the space constrained by NA\textsubscript{vis}, TE\textsubscript{0}-like slow-light resonance presents an angular tolerance higher than that of TM\textsubscript{0}-like slow-light resonance. As a result, visible excitation coupling is more efficient with TE\textsubscript{0}-like slow-light resonance than TM\textsubscript{0}-like slow-light resonance. Figures \ref{fig:BFP}e - \ref{fig:BFP}g show the overlap between the reflectance spectra of PhCs averaged over NA\textsubscript{vis} and the normalized absorption cross-section of \textsuperscript{3}F\textsubscript{4} $\rightarrow$ \textsuperscript{1}D\textsubscript{2} transition in Tm\textsuperscript{3+} denoted as $\sigma_{Tm^{3+}}$. The TM\textsubscript{0}-like slow-light resonance of PhC B overlaps with visible excitation and $\sigma_{Tm^{3+}}$. In contrast, TE\textsubscript{0}-like slow-light resonance is involved in PhC D. PhC slab C stands in the in between situation: none of its slow-light resonances in the $\Gamma$-point efficiently overlap with the visible excitation and $\sigma_{Tm^{3+}}$. Based on these observations, we argue that for the first (A) and second (D) peaks in Figure \ref{fig:raw_UC}a, the high EFs are related, respectively, to efficient excitations of the TM\textsubscript{0}-like and TE\textsubscript{0}-like slow-light resonances by visible light. The UCL dip (C) is then explained by the excited spectral range of $\sigma_{Tm^{3+}}$ standing in the gap between its two slow-light resonances. In order to extend this interpretation to all PhC slabs, we introduce the overlap factor in Equation \ref{eq:overlap_factor}:
\begin{equation}
    \gamma = \int\rho(\lambda)\times\sigma_{Tm^{3+}}(\lambda)\times R^{PhC}(\lambda)d\lambda
    \label{eq:overlap_factor}
\end{equation}
with $\rho$ the laser excitation spectral density (Figure S1-D.2b) and $R^{PhC}$ the reflectance spectra averaged on the visible numerical aperture of the considered PhC slab. As reflectance measurement is only a qualitative indicator of LDOS, due to the Fano nature of the reflectance resonances, this approach is limited to a qualitative interpretation of the impact of the overlap factor. Figure \ref{fig:FOM} presents the EF under visible and NIR excitation depending on the calculated overlap factor. The following tendency is observed: the higher the overlap factor, the higher the UV emission enhancement. This confirms the key contribution of visible slow-light resonances to the global enhancement of UCL.

%%%%    Detailed contribution of both visible and NIR modes to the overall enhancement factor

Thus, by scanning the lattice parameter, the overlap factor $\gamma$ has been finely tuned. Its highest value is associated with the strongest UCL, the PhC structure D. Its UV emission is measured to be 28.1 $\pm$ 3.0 times higher than that of the reference under both NIR and visible excitation. The extraction enhancement factor is measured to be 2.71 $\pm$ 0.5 under NIR excitation. Then, the contribution of the visible slow-light slow-light resonance is deduced to be an enhancement of the ESA by a factor of 10.4 $\pm$ 2.2. As only one visible photon is involved in the multi-wavelength upconversion mechanism (Figure \ref{fig:UC_process}), we deduce that the visible slow-light resonance enhanced the local excitation irradiance by a factor of 10.4 $\pm$ 2.2.
\FloatBarrier

\section*{Conclusions}
%%%%    Conclusion

In conclusion, we demonstrated a design free from parasitic absorption losses, which enhances UV emission of the Yb\textsuperscript{3+}-Tm\textsuperscript{3+} co-doped system through multi-wavelength excitation in the NIR and the visible ranges by a factor of 28.1 $\pm$ 3.0. To the best of our knowledge, this is the first demonstration of multi-wavelength excitation assisted by LDOS engineering. Contributions of visible light absorption and UV light extraction enhancements to the overall enhancement of UCL were measured to be, respectively, factors of 10.4 $\pm$ 2.2 and 2.71 $\pm$ 0.5. This upconversion mechanism involving one NIR photon and one visible photon was confirmed through power law characterization. The properties of Bloch modes were tuned by scanning the lattice parameter of PhC structures, and the measured band structures are consistent with the simulations. The key role of slow-light resonance in overcoming the limited absorption cross-section in lanthanides, and thus in enhancing the visible ESA, was highlighted.

%%%%    Outlooks

The methodology developed in this paper to simulate, fabricate, and characterize PhC structures that enhance the upconversion mechanism involving ESA in lanthanides could be applied to other transitions or systems. For example, fluorinated host matrices are also interesting because they benefit from considerable upconverting efficiencies \cite{NatNano_Zhou_2015, ACS_Purohit_2019}, and additional sensitizers and activators could be considered to address even wider excitation spectral ranges \cite{Crystals_Vu_2021, Nat_Schiattarella_2024, ACS_Tseng_2024, ChemComm_Yeow_2025}. In addition, the demonstrated design enhances the multi-wavelength upconversion mechanism that governs UV emission of Yb\textsuperscript{3+}-Tm\textsuperscript{3+} under solar excitation. They could be fabricated using large scale lithography techniques, such as nanoimprint \cite{ACS_Würth_2020} or laser interference \cite{ACS_Hoang_2017}, thus paving the way toward solar-based applications. 

Enhancement of ESA in lanthanides with PhC would also be of great interest for photon-avalanching regimes, by reducing their thresholds and increasing their non-linearities \cite{ACS_Le-Vu_2024}.

\section*{Methods}
\noindent\textbf{Synthesis of Y\textsubscript{2}O\textsubscript{3}:Yb\textsuperscript{3+},Yb\textsuperscript{3+} Target for PLD.} Powder precursors of Y\textsubscript{2}O\textsubscript{3} (Sigma-Aldrich - Ref. 205168), Yb\textsubscript{2}O\textsubscript{3} (Sigma-Aldrich - Ref. 204676) and Tm\textsubscript{2}O\textsubscript{3} (Sigma-Aldrich - Ref. 637300) were mixed for one hour at 50 rpm using a 3D Shaker Mixer Type M10 from Beijing Grinder Instrument, then pressed at 10T using the Manual Hydraulic Press from Specac. The target obtained was annealed at 1400°C for 20h with an increase and decrease in temperature of 3°C / min. The target doping rates were optimized to maximize the \textsuperscript{1}D\textsubscript{2} $\rightarrow$ \textsuperscript{3}H\textsubscript{6} UV emission. Indeed, a trade-off was found between increases in energy transfer efficiency from Yb\textsuperscript{3+} to Tm\textsuperscript{3+} and in UV emitter concentration, and the limitation of the doping quenching effect. The optimized doping rates are the following: 7.5\% of Yb\textsuperscript{3+} and 0.5\% of Tm\textsuperscript{3+}. Additional x-ray diffraction and luminescence characterization of the target are provided in the Supplemental Information \ref{app:target_PLD}.

\noindent\textbf{Deposition of the Y\textsubscript{2}O\textsubscript{3}:Yb\textsuperscript{3+},Yb\textsuperscript{3+} Thin Film.} The thin film of Y\textsubscript{2}O\textsubscript{3}:Yb\textsuperscript{3+},Tm\textsuperscript{3+} was deposited by PLD at room temperature under an oxygen pressure of P\textsubscript{O\textsubscript{2}} = 10\textsuperscript{-4} mbar. We used a home-made target and a top-hat KrF excimer laser (Coherent COMPex 201). The latter delivers 20 ns pulses at a repetition rate of 5 Hz and a wavelength of 248 nm. The target was ablated at a fluence of F = 2.23 $\pm$ 0.28 J/cm\textsuperscript{2} for 12.5 minutes with a spot size of 3.93 $\pm$ 0.45  mm\textsuperscript{2}. After an annealing at 600°C for 6h with an increase and decrease in temperature of 0.5°C / min, the thickness of the thin film was measured with a profilometer to be in the range 70 - 95 nm on a surface of 2 cm $\times$ 2 cm. This non-uniformity comes from the angular distribution of the plasma in the pulsed laser deposition chamber \cite{PLD_Chrisey_1994}. Its roughness was measured to be 2.8 $\pm$ 0.1 nm with an atomic force microscope. Additional information on the host matrix thin film dispersion is provided in the Supplemental Information \ref{app:PLD_thinfilms}.

\noindent\textbf{Deposition of the SiN\textsubscript{x} Thin Film.} The thin film of SiN\textsubscript{x} was deposited by PECVD (Advanced Vacuum Vision 310) at 300°C and 600 mTorrs, using a gas mixture of silane (10 sccm), ammonia (20 sccm) and nitrogen (600 sccm), working at 30 W. Following a deposition of 17 minutes, the thickness of the thin film was measured by ellipsometry to be 144.0 $\pm$ 0.6 nm. The overall roughness of the heterogeneous stack was measured to be 3.2 $\pm$ 1.0 nm by atomic force microscopy. Additional information on the thin film dispersion is provided in the Supplemental Information \ref{app:PECVD_thinfilms}.

\noindent\textbf{RCWA Simulations.} The band structures and absorption responses of the PhCs were calculated using the Python API of the RCWA simulation tool "S\textsuperscript{4}" developed by the group of S. Fan \cite{ComPhysComm_Liu_2012}. Considerations such as material dispersions, excitations, the scanning of the space of parameter, as well as our figure of merit, are detailed in the Supplemental Information \ref{app:RCWA}.

\noindent\textbf{Electron-beam Lithography.} PhC patterns were defined by electron-beam lithography on 60 µm $\times$ 60 µm areas. The setup consists in an Inspect F FEI scanning electron microscope (30 keV), equipped with a laser alignment stage connected to an Elphy Plus External module operated by the Raith lithography software. A 200 nm thick positive resist, AR-P 6200.09, was spin-coated for 60s at 4000 rpm with a 3000 s\textsuperscript{-2} acceleration, then post-backed at 150°C for 60s. Then, an additional conductive coating, AR-PC 5092.02, was spin-coated for 60s at 4000 rpm with a 1000 s\textsuperscript{-2} acceleration to avoid charging effect during the lithography process. It was post-backed at 90°C for 120s. The electron-beam lithography exposure dose was 80 µC/cm\textsuperscript{2}. The protective coating was rinsed in two deionized water baths for 60s each. Then the positive resist was developed at room temperature with two successive baths under manual agitation: developer AR-P 600-546 for 60s, then stopper AR-P 600-60 for 60s.

\noindent\textbf{Etching of SiN\textsubscript{x}.} The patterns were transferred in SiN\textsubscript{x} by ICP-RIE (Sentech SI 500) at 1.6 Pa and 20°C, with a gas mixture of CHF\textsubscript{3} (25 sccm) and Ar (3 sccm). The chamber was first passivated, and then the sample was etched by 30-second etching and pumping sequences at a bias of -81 V. The remaining resist mask is removed with an oxygen plasma at 2.0 Pa, 25°C, 50 sccm of O\textsubscript{2} and a bias of -48 V. We obtained homogeneous patterns with a filling factor of 47.9\% $\pm$ 0.2\% and an etching depth of 138 $\pm$ 1 nm according to the SEM top-view image and the profile of patterns measured by atomic force microscopy presented in Figures \ref{fig:design_illustration}b \& \ref{fig:AFM_profil}.

\noindent\textbf{UCL Measurement.} A presentation of our experimental setup is provided in Supplemental Information \ref{app:UCL_setup}. It includes a scheme of the setup and the details of our sources, optics, and detector. Spectra were acquired by averaging 10 measures that were integrated for 6s. The background signal was subtracted for each measurement.

\noindent\textbf{Band structure Measurement.} A presentation of our experimental setup is provided in Supplemental Information \ref{app:setup_reflectivity}. It includes a scheme of the setup and the details of our sources, optics, and detector, as well as the method to measure the reflectivity of a sample.

\section*{Acknowledgments}
\noindent The authors acknowledge funding from the Agence Nationale de la Recherche through grant number ANR-22-CE09-0036, the French region Rhône Alpes Auvergne and the Grand Lyon metropole (CPER SULTRANSE). 
The authors acknowledge support from the CNRS/IN2P3 Computing Center (Lyon - France) and HPC resources of PMCS2I (Pôle de Modélisation et de Calcul en Sciences de l'Ingénieur de l'Information) of the École Centrale de Lyon (Écully - France) for providing computing and data-processing resources needed for this work.
The authors acknowledge support from Nanolyon platform, a member of the CNRS-RENATECH+ French national network, for providing access to clean-room and characterization facilities.

\printbibliography[title={References}]

\end{refsection}

\begin{refsection}

\onecolumn

\begin{center}
    \centering
    \Huge Supplemental Information
\end{center}

\vspace{0.5cm}

\appendix

\section{UV upconversion mechanisms}
\subsection{NIR and NIR plus visible excitations}
\label{app:UC}

\begin{figure}[!h]
    \centering
    \includegraphics[]{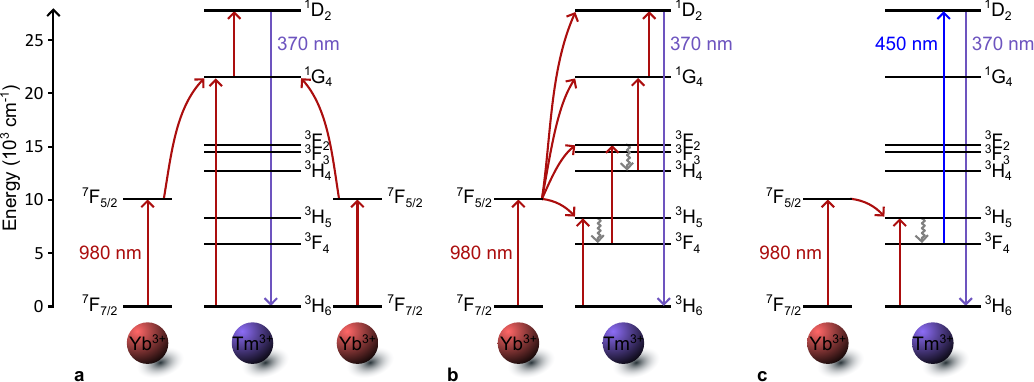}
    \caption{UV Upconversion Mechanisms in Yb\textsuperscript{3+}-Tm\textsuperscript{3+} co-doped system. (a) Cooperative energy transfer upconversion. (b) Sequential energy transfer upconversion. (c) Multi-wavelength mechanism, involving energy transfer upconversion and excited-state absorption.}
    \label{fig:UC_process_app}
\end{figure}

We consider the co-doped system Yb\textsuperscript{3+}-Tm\textsuperscript{3+} under NIR ($\sim$ 980 nm) excitation. UV emission is obtained through cooperative energy transfer upconversion (Figure \ref{fig:UC_process_app}.a) and sequential energy transfer upconversion (Figure \ref{fig:UC_process_app}.b), involving, respectively, three and four photon absorptions \cite{NatMat_Wang_2011, RSC_Mishra_2012, ACS_Purohit_2019}. These two mechanisms take advantage of the ladder-like repartition of the energy diagram of thulium ions. Now, considering an additional excitation in the visible range ($\sim$ 450 nm), UV emission is obtained through the multi-wavelength mechanism presented in Figure \ref{fig:UC_process_app}.c and detailed in the article.

\section{Simulations}
\subsection{Rigorous coupled-wave analysis - Excited-state absorption enhancement}
\label{app:RCWA}

The definition of media dispersion ($\tilde{n} = n + i\kappa$) is of major importance in electromagnetic simulations. These of SiO\textsubscript{2} and SiN\textsubscript{x} were experimentally measured by ellipsometry (Appendix \ref{app:fabrication}). However, the imaginary part of Y\textsubscript{2}O\textsubscript{3}:Yb\textsuperscript{3+},Tm\textsuperscript{3+} cannot be fully assessed by classical ellipsometry measurement, since transitions such as \textsuperscript{3}F\textsubscript{4} $\rightarrow$ \textsuperscript{1}D\textsubscript{2} are excited-state absorptions. Thus, the latter extinction coefficient, $\kappa_{{}^3\!F_4\rightarrow{}^1\!D_2}$, was estimated based on the excitation spectrum of Y\textsubscript{2}O\textsubscript{3}:Yb\textsuperscript{3+},Tm\textsuperscript{3+} (Supplemental Information S1-C2), the Judd-Ofelt theory \cite{JoL_Hehlen_2013} and assuming that \textsuperscript{3}F\textsubscript{4} was highly populated under NIR excitation:
$$\tilde{n}_{Y_2O_3:Yb^{3+},Tm^{3+}}=\tilde{n}_{Y_2O_3} + i\kappa_{{}^3\!F_4\rightarrow{}^1\!D_2}$$
where $\tilde{n}_{Y_2O_3}$ was measured by ellipsometry on a thin non-doped film of yttrium oxide.

\begin{figure}[t]
    \centering
    \includegraphics[]{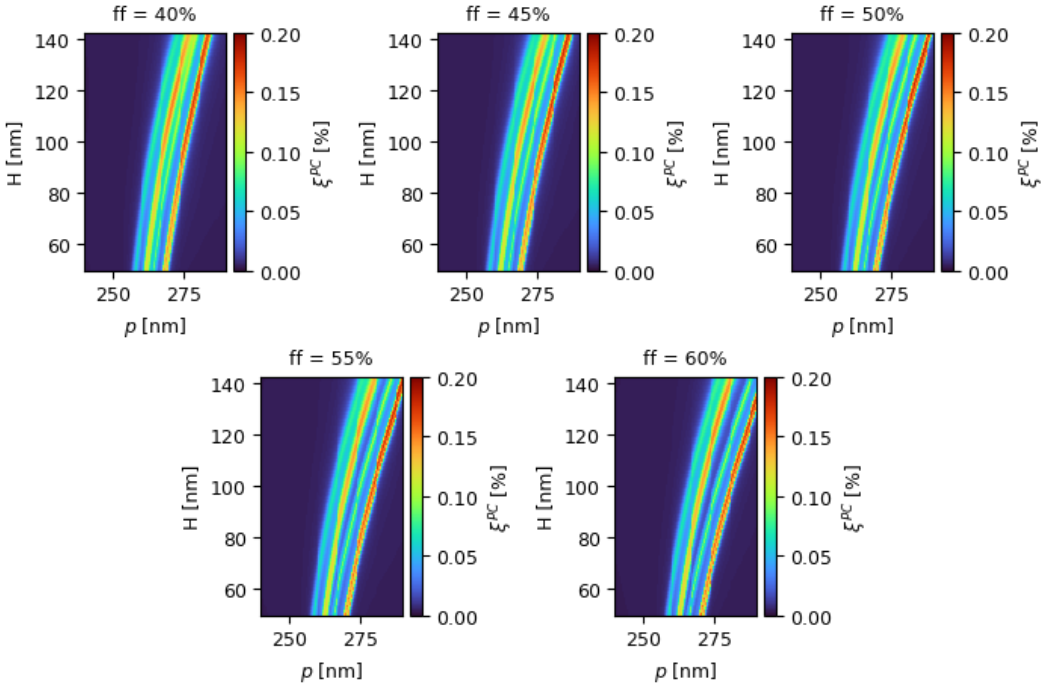}
    \caption{Maps of the Figure of Merit obtained from Rigorous Coupled-Wave Analysis simulations by scanning geometric characteristics of the photonic structure (p, H, ff). Harmonics were retained at 49.}
    \label{fig:RCWA_scan}
\end{figure}

\medskip

Experimentally, a $\sim$ 144 nm thick SiN\textsubscript{x} thin film was deposited on top of a $\sim$ 95 nm thick Y\textsubscript{2}O\textsubscript{3}:Yb\textsuperscript{3+},Tm\textsuperscript{3+} thin film deposited on top of a fused silica wafer. To tune a slow-light resonance in the $\Gamma$-point with the excited-state absorption \textsuperscript{3}F\textsubscript{4} $\rightarrow$ \textsuperscript{1}D\textsubscript{2} and thus increase the number of photons absorbed from an unpolarized light source, we considered both transverse electric and transverse magnetic excitations, and scanned the geometric characteristics of the pattern: its lattice parameter (p), depth (H) and air filling factor (ff), which also depends on the radius (r) of the pattern ($ff=\pi r^2 / p^2$). We defined the figure of merit, $\xi$, which takes into account the sun as the light source and is presented in Equation \ref{eq:sol_abs} \cite{ASTM_2020}. Involved terms are the following:

\begin{itemize}
    \item The spectral distribution of the number of photons per unit of time that irradiate the sample at normal incidence: $n_{ph}$.
    \item The absorption response of the design that depends on its geometric characteristics: $A$.
\end{itemize}

\begin{equation}
    \xi^{(i)} = \frac{\int{n_{ph}(\lambda)A^{(i)}(\lambda)d\lambda}}{\int{n_{ph}(\lambda)d\lambda}}
    \label{eq:sol_abs}
\end{equation}

\medskip

To determine the initial lattice parameter of the photonic structure, we solved the modes in the unpatterned stack to obtain the propagating constant of the TE\textsubscript{0} guided mode (labeled $\beta$) \cite{Solver_Hammer_2025}, and we considered $p\simeq 2\pi/\beta$. Maps of $\xi$ are presented in Figure \ref{fig:RCWA_scan} for p $\in$ [240 ; 290] nm, H $\in$ [50 ; 144] nm, ff $\in$ [40 ; 60] \%. They enabled us to determine what designs of photonic crystals were worth to fabricate.

\section{Fabrication}
\label{app:fabrication}
PhC fabrication was performed with a top-down approach. It consisted of the deposition of thin films of Y\textsubscript{2}O\textsubscript{3}:Yb\textsuperscript{3+},Tm\textsuperscript{3+} and SiN\textsubscript{x}, followed by the patterning of SiN\textsubscript{x}. We used a fused silica wafer from BT Electronics, which is transparent from the NIR absorption to the UV emission of the lanthanide-doped system of interest, and has a lower refractive index than our thin films. This appendix contains additional fabrication information to that given in the Method section of the article.

\subsection{Characterization of materials}
\label{app:charac_mat}

\subsubsection{Y\textsubscript{2}O\textsubscript{3}:Yb\textsuperscript{3+}Tm\textsuperscript{3+} - Pulsed Laser Deposition Target}
\label{app:target_PLD}

\begin{figure}[!h]
    \centering
    \includegraphics[]{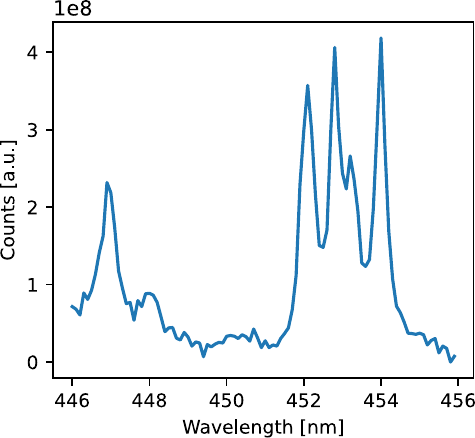}
    \caption{Shape of the \textsuperscript{3}F\textsubscript{4}$\rightarrow$\textsuperscript{1}D\textsubscript{2} excited-state absorption determined by excitation measurement of the pulsed laser deposition target of Y\textsubscript{2}O\textsubscript{3}:Yb\textsuperscript{3+}(7.5\%),Tm\textsuperscript{3+}(0.5\%) while pumping the system at 973 nm.}
    \label{fig:excitation_spectrum}
\end{figure}

The knowledge of the spectral shape of the excited-state absorption transition \textsuperscript{3}F\textsubscript{4}$\rightarrow$\textsuperscript{1}D\textsubscript{2} is of major interest for engineering the local photonic density of state of photonic crystals. Hence, we measured the UV emission at 367 nm of the pulsed laser deposition target of Y\textsubscript{2}O\textsubscript{3}:Yb\textsuperscript{3+}(7.5\%),Tm\textsuperscript{3+}(0.5\%) as we scanned the wavelength of an optical parametric oscillator (EKSPLA - NT230-50-SH) excitation while pumping the material with a continuous-wave laser at 973 nm to populate the excited-state \textsuperscript{3}F\textsubscript{4}. The resulting excitation spectrum of the transition is presented in Figure \ref{fig:excitation_spectrum}.

\medskip

\begin{figure}[!h]
    \centering
    \includegraphics[]{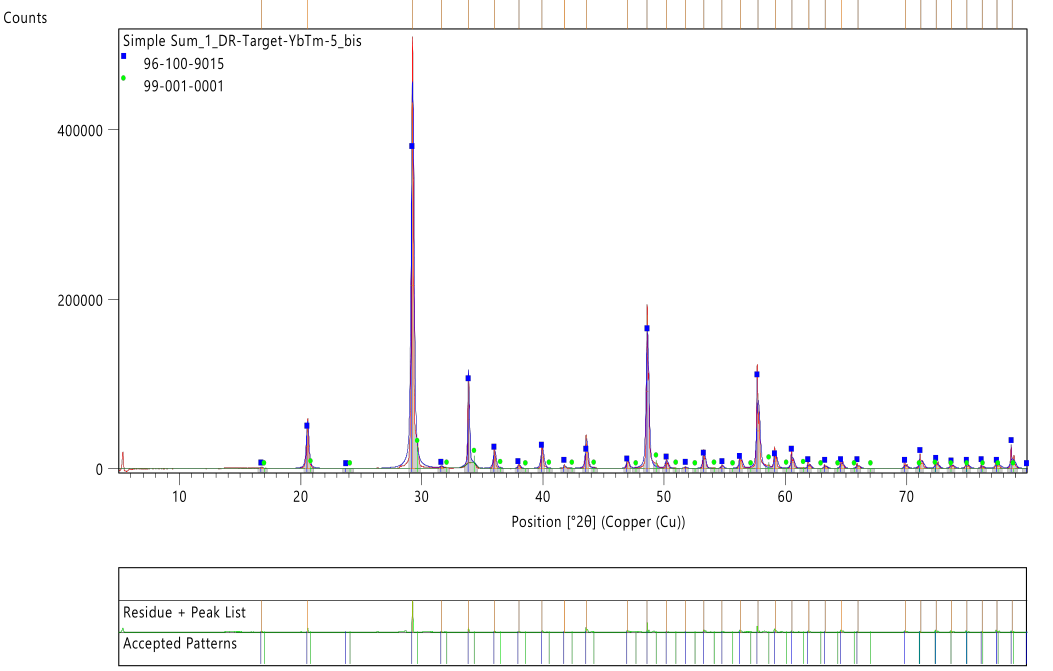}
    \caption{XRD measurement of the pulsed laser deposition target of Y\textsubscript{2}O\textsubscript{3}:Yb\textsuperscript{3+}(7.5\%),Tm\textsuperscript{3+}(0.5\%). The red, blue and green curves respectively correspond to the experimental measurement of Y\textsubscript{2}O\textsubscript{3}:Yb\textsuperscript{3+}(7.5\%)Tm\textsuperscript{3+}(0.5\%) host matrix, and the host matrices data extracted from the literature of Y\textsubscript{2}O\textsubscript{3} and Yb\textsubscript{2}O\textsubscript{3} \cite{MatSciFor_Baldinozzi_1998, JAC_Farhat_2009}.}
    \label{fig:Target_XRD}
\end{figure}

X-ray diffraction (XRD) was performed to measure the crystalline phase of the target, using the Aeris device from Malvern Panalytical. The signal obtained from Y\textsubscript{2}O\textsubscript{3}:Yb\textsuperscript{3+}(7.5\%)Tm\textsuperscript{3+}(0.5\%) is shown in Figure \ref{fig:Target_XRD}. It presents a positive offset of $\sim$ 0.056° compared to that of Y\textsubscript{2}O\textsubscript{3} \cite{MatSciFor_Baldinozzi_1998}. It can be explained by the Yb\textsuperscript{3+} doping, which is dominant compared to the Tm\textsuperscript{3+} doping. Indeed, the XRD signal of the matrix of Yb\textsubscript{2}O\textsubscript{3} is shifted by 0.457° compared to that of Y\textsubscript{2}O\textsubscript{3} \cite{JAC_Farhat_2009}. Thus, we can confirm the cubic phase of the target of Y\textsubscript{2}O\textsubscript{3}:Yb\textsuperscript{3+}(7.5\%)Tm\textsuperscript{3+}(0.5\%). As trivalent lanthanide ions substitute the yttrium in the host matrix, we can derive the activator density to be N\textsubscript{Tm\textsuperscript{3+}} = 1.34 $\times$ 10\textsuperscript{8} µm\textsuperscript{-3}.

\subsubsection{Y\textsubscript{2}O\textsubscript{3} - Thin film dispersion}
\label{app:PLD_thinfilms}

\begin{figure}[!h]
    \centering
    \includegraphics[]{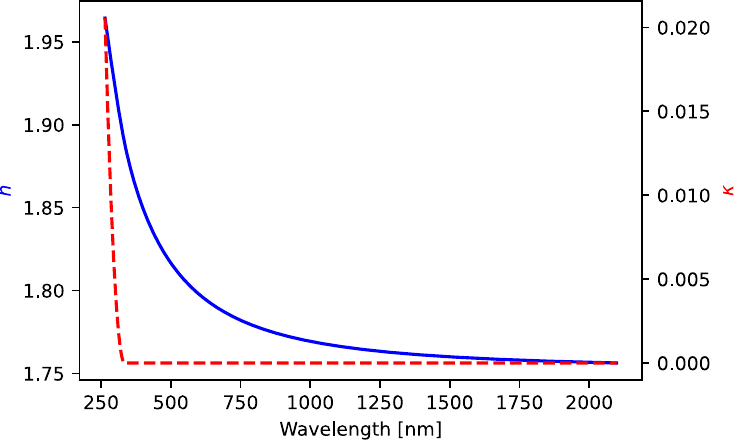}
    \caption{Dispersion of the Y\textsubscript{2}O\textsubscript{3} host matrix thin film (62.6 $\pm$ 0.1 nm)}
    \label{fig:Y2O3_HM-Disp}
\end{figure}

The dispersion of the host matrix $\tilde{n}_{Y_2O_3}$ was measured with the M-2000 Spectroscopic Ellipsometer (J.A. Woollam) by fitting the data with the Tauc-Lorentz model. The fitting parameter obtained was $\chi^2$ = 1.37. It is presented in Figure \ref{fig:Y2O3_HM-Disp}.

\subsubsection{SiN\textsubscript{x} - Thin film dispersion}
\label{app:PECVD_thinfilms}

\begin{figure}[!h]
    \centering
    \includegraphics[]{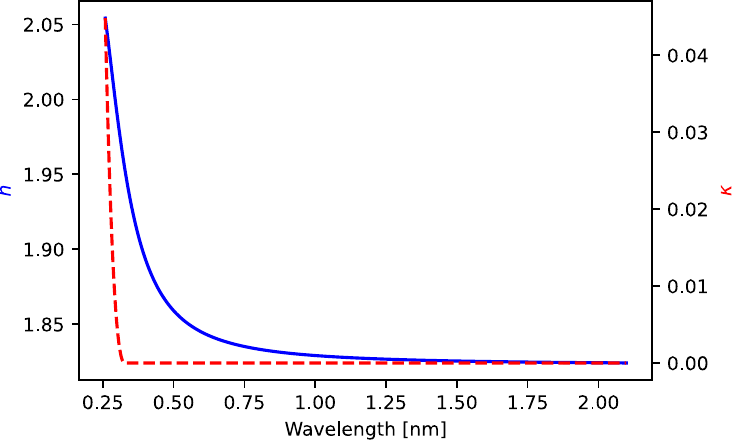}
    \caption{Dispersion of the SiN\textsubscript{x} thin film (144 $\pm$ 0.6 nm)}
    \label{fig:SiN-Disp}
\end{figure}

The dispersion of the optimized thin film $\tilde{n}_{SiN_x}$ was measured with the M-2000 Spectroscopic Ellipsometer (J.A. Woollam) by fitting the data with the New Amorphous model. The fitting parameter obtained was $\chi^2$ = 2.50. It is presented in Figure \ref{fig:SiN-Disp}. Particular attention was paid to the SiH\textsubscript{4}/NH\textsubscript{3} flux ratio as the SiN\textsubscript{x} energy gap and refractive index increase with the SiH\textsubscript{4}/NH\textsubscript{3} ratio. A trade-off was found to enable transparency from NIR absorption to UV emission of the lanthanide-doped system while preserving a refractive index as high as possible.

\subsection{SiN\textsubscript{x} etching}
\label{app:etching}

\begin{figure}[!h]
    \centering
    \includegraphics[]{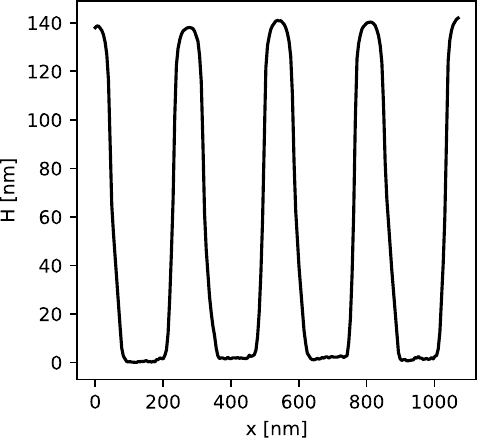}
    \caption{Atomic force microscopy profile of the etching depth of the photonic structure D of the article.}
    \label{fig:AFM_profil}
\end{figure}

The etching depth of the PhC patterns was characterized with an atomic force microscope. The profile measure presented in Figure \ref{fig:AFM_profil} is the result from the convolution of the microscope tip and the PhC surface.

\FloatBarrier

\vfill
\section{Characterization}
\subsection{Upconversion luminescence measurement setup}
\label{app:UCL_setup}

\begin{figure}[t]
    \centering
    \includegraphics[]{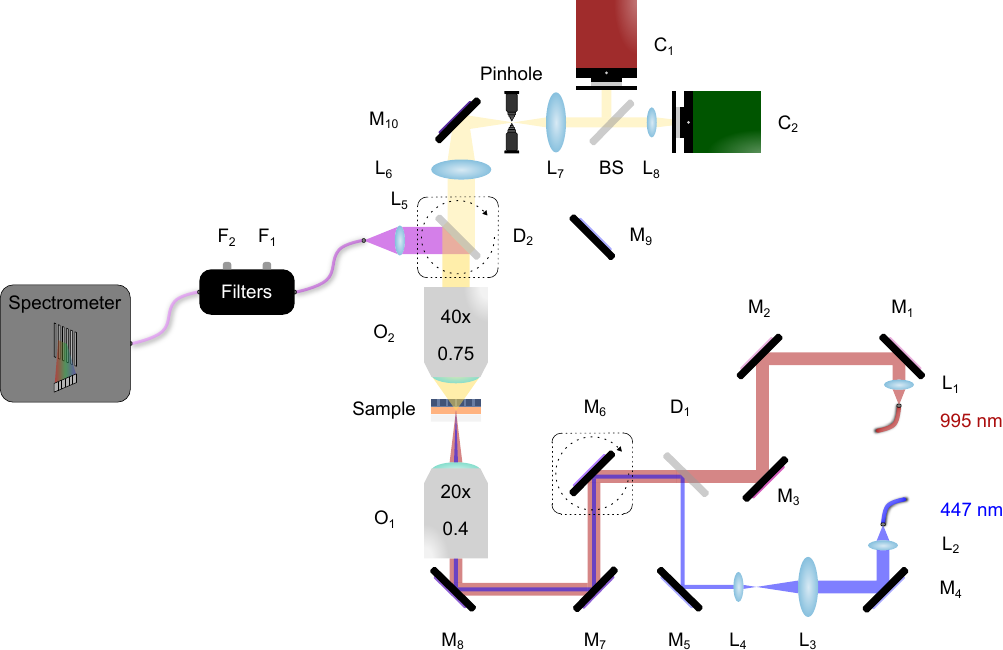}
    \caption{Optical setup used for measurements of the upconversion luminescence presented in the transmission configuration}
    \label{fig:micro-uc}
\end{figure}

The setup used for the measurements of the upconversion luminescence is presented in Figure \ref{fig:micro-uc}. References of the components used for the setup are given in Table \ref{tab:ref_micro-uc}.

\medskip

\begin{figure}[b]
    \centering
    \includegraphics[]{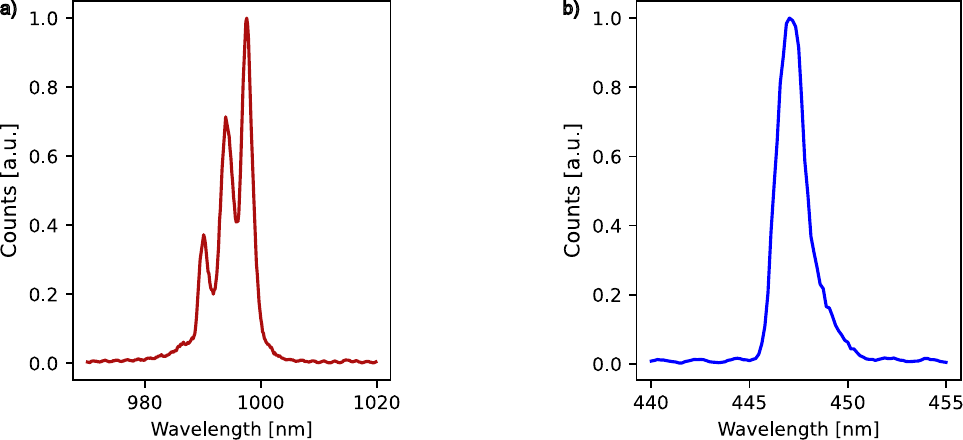}
    \caption{Emission spectra of the laser used for upconversion luminescence measurements are respectively presented in (a) for L\textsubscript{1} in the IR and (b) for L\textsubscript{2} in the visible.}
    \label{fig:laser_em_spectra}
\end{figure}

Mirrors M\textsubscript{6} and D\textsubscript{2} can rotate to switch from transmission to reflection configuration. In this paper, we used the setup in the transmission configuration. The position of the sample holder can be fine-tuned in all directions ($\vec{e}_x$, $\vec{e}_y$ and $\vec{e}_z$). Continuous-wave (CW) lasers with Gaussian beams are collimated to excite the sample through the microscope objective O\textsubscript{1}. The emission spectra of our laser are presented in Figure \ref{fig:laser_em_spectra}. The numerical aperture of excitation in the NIR is NA\textsubscript{NIR} = 0.4, while the size of the visible laser is reduced with the telescope made of L\textsubscript{3} and L\textsubscript{4} to reduce its effective numerical aperture of excitation to NA\textsubscript{vis} = 0.04. The UV signal is collected through the high numerical aperture microscope objective O\textsubscript{2} (NA\textsubscript{UV} = 0.75). The dichroic mirror D\textsubscript{2} reflects UV towards the measurement line and transmits the visible and NIR towards the imaging line. The measurement line is composed of filters to remove residual transmitted excitation, and a spectrometer. The imaging line enables us to both image the back focal plane (C\textsubscript{1}) to characterize the direction of emission and transmission, and the sample (C\textsubscript{2}) to align the latter with the excitations. The pinhole enables us to limit ambient parasitic photons and adjust the studied area of the sample.

\begin{table}[!h]
\centering
\begin{tabular}[t]{|c|l|}
\hline
\multicolumn{1}{|c|}{\textbf{Label(s)}} & \multicolumn{1}{|c|}{\textbf{Reference / Characteristics}} \\ \hline
\multicolumn{1}{|c|}{CW laser @995 nm}         & \multicolumn{1}{|l|}{Integrated Optics - 980 nm laser}          \\ \hline
\multicolumn{1}{|c|}{CW laser @447 nm}         & \multicolumn{1}{|l|}{Integrated Optics - 450 nm laser}          \\\hline
 M\textsubscript{1}, M\textsubscript{2}&R>99\% @800-1000 nm\\\hline 
                               M\textsubscript{3}&                                Thorlabs - BB1-E03\\ \hline
                               M\textsubscript{4}, M\textsubscript{5}, M\textsubscript{9}&                                Thorlabs - BB1-E02\\\hline
 M\textsubscript{6}&Thorlabs - MRA25-P01\\\hline
 M\textsubscript{7}, M\textsubscript{8}&Thorlabs - PF10-03-P01\\\hline
 M\textsubscript{10}&Thorlabs - PF20-03-F01\\\hline
                               D\textsubscript{1}&                                Thorlabs - DMLP650R\\ \hline
                               D\textsubscript{2}&                                Thorlabs - DMLP425R\\ \hline
                               BS&                                10\% towards C\textsubscript{1} \& 90\% towards C\textsubscript{2}\\ \hline
                               O\textsubscript{1}&                                Mitutoyo -  M PLAN NIR 20x | NA = 0.4\\ \hline
                               O\textsubscript{2}&                                Nikon -  Plan Fluor 40x | NA = 0.75\\ \hline
                               L\textsubscript{1}&                                Thorlabs - AL1225H-B\\ \hline
                               L\textsubscript{2}&                                Thorlabs - AL1225G-A\\ \hline
                               L\textsubscript{3}&                                Thorlabs - LA1509-ML\\ \hline
                               L\textsubscript{4}&                                Thorlabs - LA1540-ML\\\hline
 L\textsubscript{5}&Thorlabs - LA4306-ML\\ \hline
                               L\textsubscript{6}&                                f' = 200 mm\\ \hline
                               L\textsubscript{7}&                                Thorlabs - AC508-080-AB-ML\\ \hline
                               L\textsubscript{8}&                                Thorlabs - AC254-040-A-ML\\ \hline
                               F\textsubscript{1}&                                Andover Corporation - KRON-U-25\\ \hline
                               F\textsubscript{2}&                                HOYA - U330\\ \hline
                               C\textsubscript{1}&                                Allied Vision - Alvium 1800 U-501 NIR\\ \hline
                               C\textsubscript{2}&                                JAI - GO-2400M-USB\\ \hline
                               Spectrometer&                                Kymera 328i Specropraph equiped with DU970P-UVB EMCCD\\ \hline
\end{tabular}
\caption{References of the components used for the setup to measure the upconversion luminescence.}
\label{tab:ref_micro-uc}
\end{table}

\subsection{Power law fit}
\label{app:powerlaw}

The dependence of the upconversion luminescence irradiance on the excitation irradiance in the NIR and in the visible is described in Equations \ref{eq:I_UCL} \& \ref{eq:I_UCL_log}. Involved terms are the following:

\begin{itemize}
    \item Upconversion luminescence irradiance: $I_{UCL}$
    \item Local NIR excitation irradiance: $I_{NIR}^{loc}$
    \item Local visible excitation irradiance: $I_{vis}^{loc}$
    \item NIR non-linear dependency: $n_{NIR}$
    \item Visible non-linear dependency: $n_{vis}$
\end{itemize}

\begin{equation}
    I_{UCL} \propto \left(I_{NIR}^{loc}\right)^{n_{NIR}} \times \left(I_{vis}^{loc}\right)^{n_{vis}}
    \label{eq:I_UCL-app}
\end{equation}

\begin{equation}
    \log(I_{UCL}) = \log(C) + n_{NIR} \times \log(I_{NIR}^{loc}/I_0) + n_{vis} \times \log(I_{vis}^{loc}/I_0)
    \label{eq:I_UCL_log}
\end{equation}

\medskip

To fit the experimental data, we processed with the logarithmic expression presented in Equation \ref{eq:I_UCL_log}, where $I_0$ is a homogeneity constant and C a dimensionless constant. Thus, non-linear orders $n_{NIR}$ and $n_{vis}$ were experimentally determined by performing, respectively, the following linear regressions: $\log(I_{UCL})$ as a function of $\log(I_{NIR}^{loc}/I_0)$ and $\log(I_{UCL})$ as a function of $\log(I_{vis}^{loc}/I_0)$. To do so, the \textit{stats.linregress} function from \textit{scipy} package (1.16.0 version) in \textit{Python} was used. Table \ref{tab:lin_reg} presents the resulting non-linear orders and correlation factors.

\begin{table}[!h]
\centering
\begin{tabular}{c|c|}
 \textbf{Non-linear order}& \textbf{Correlation factor ($r^2$)} \\ \hline
 $n_{NIR} = 1.0$& 0.997                                   \\ \hline
 $n_{vis} = 1.0$& 0.998                                   \\ \hline
\end{tabular}
\caption{Non-linear orders determined by linear regression of experimental data}
\label{tab:lin_reg}
\end{table}

\vfill
\subsection{Response of the photonic crystal structure to a linearly polarized visible excitation}
\label{app:pola}

In order to experimentally characterize the response of the PhC D, we measured UCL spectra under the multi-wavelength excitation regime with different linear polarizations of the visible excitation. As the visible laser is not perfectly linearly polarized, we set a polarizer (Thorlabs - WP25M-VIS) at the output of L\textsubscript{4} (see Figure \ref{fig:micro-uc} and Table \ref{tab:ref_micro-uc}) and measured the power at its output for each polarization considered.

\medskip

\begin{figure}
    \centering
    \includegraphics[]{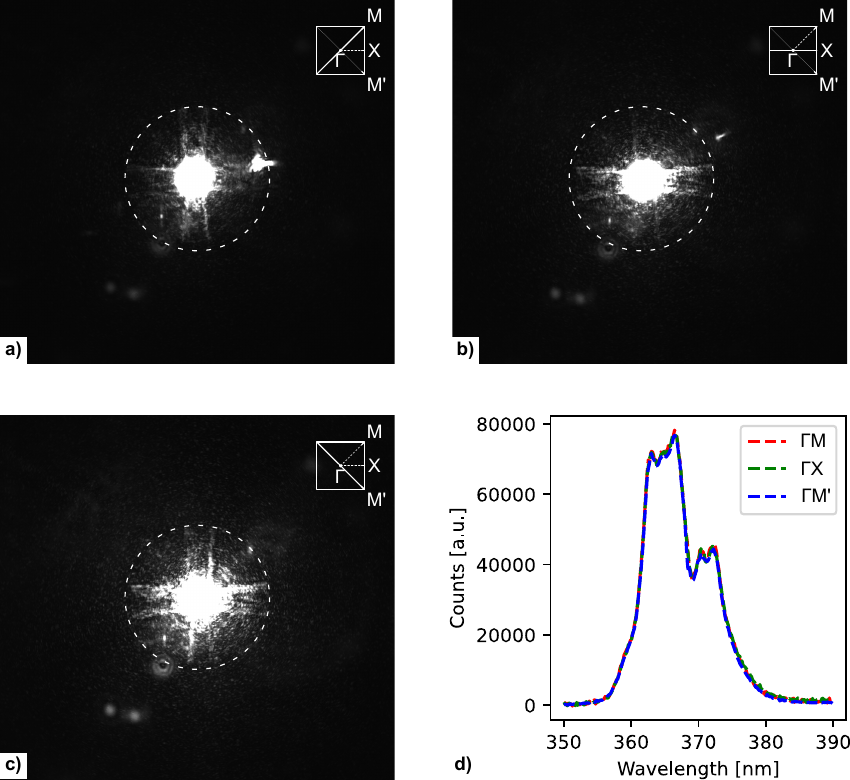}
    \caption{Characterization of the photonic crystal D under the multi-wavelength regime with the visible excitation polarized along $\Gamma$M, $\Gamma$X and $\Gamma$M' high symmetry directions. (a), (b), and (c) Fourier imaging of the blue emission, respectively for the polarization oriented along the direction $\Gamma$M, $\Gamma$X and $\Gamma$M'. (d) UV upconversion luminescence spectra.}
    \label{fig:pola}
\end{figure}

We captured the direction of  blue transmission and emission by Fourier space imaging with camera C\textsubscript{1}. In Figures \ref{fig:pola}a and \ref{fig:pola}c, we observed transmission and emission through Bloch-modes along $\Gamma$X and its perpendicular direction with the same intensity. Thus, it confirmed that the excitation polarization was along $\Gamma$M and $\Gamma$M'. In contrast, in Figure \ref{fig:pola}b, we mainly observed the signal along $\Gamma$X, which confirmed that the polarization of the visible excitation was along $\Gamma$X. UV UCL spectra were measured for these three polarizations. A correction factor was applied to each of them, taking into account the power measured at the output of the polarizer and the power law of the multi-wavelength regime, to compare the measurements in Figure \ref{fig:pola}d. The UCL obtained under the different excitation conditions are strongly similar.

\medskip

Nevertheless, these are not enough to argue that the UCL of the photonic crystal D is highly tolerant to the polarization of excitation. Indeed, the dispersion-band of the considered slow-light mode is not entirely flat in the numerical aperture considered NA\textsubscript{vis} (Figure 4), and our excitation is not exactly monochromatic (Figure \ref{fig:laser_em_spectra}b). In this respect, additional measurements for linear polarization should be performed in directions in between $\Gamma$X and $\Gamma$M.

\subsection{Reflectivity measurement setup}
\label{app:setup_reflectivity}

\begin{figure}[!h]
    \centering
    \includegraphics[]{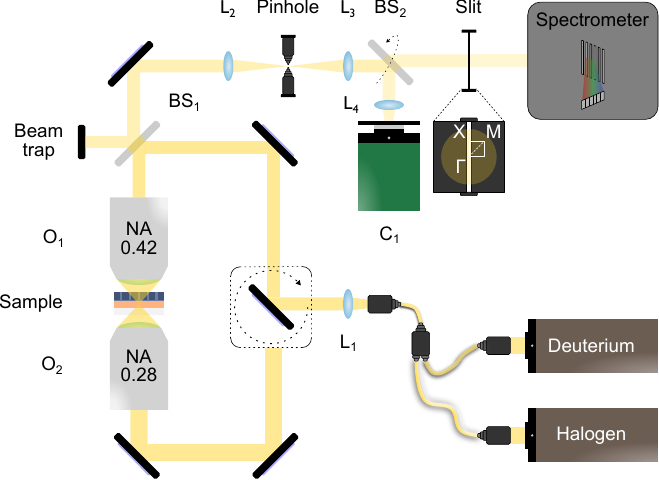}
    \caption{Optical setup used to measure band structures of photonic crystals presented in the reflection configuration}
    \label{fig:BFP-Setup}
\end{figure}

The setup used to measure the band structure of photonic crystals is presented in Figure \ref{fig:BFP-Setup}. References of the components used for the setup are given in Table \ref{tab:ref_micro-uc}.

\begin{table}[!h]
\centering
\begin{tabular}{|c|l|}
\hline
\multicolumn{1}{|c|}{\textbf{Label}} & \multicolumn{1}{|c|}{\textbf{Reference}} \\ \hline
\multicolumn{1}{|c|}{White light source}         & \multicolumn{1}{|l|}{Ocean Optics - DH-2000}          \\\hline
 L\textsubscript{1}&Thorlabs - LA4052\\\hline
 L\textsubscript{2}&Thorlabs - AC254-200-A-ML\\\hline
                               L\textsubscript{3}&                                Thorlabs - LA4380\\ \hline
                               L\textsubscript{4}&                                Thorlabs - LA1433-A\\ \hline
                               O\textsubscript{1}&                                Mitutoyo - M Plan Apo NUV 20x | NA = 0.42\\ \hline
                               O\textsubscript{2}&                                Mitutoyo - M Plan Apo 20x | NA = 0.42\\ \hline
                               BS\textsubscript{1}&                                Thorlabs - BSW27\\ \hline
                               BS\textsubscript{2}&                                Thorlabs - CCM1-BS013/M\\ \hline
                               Spectrometer&                                Horiba - MicroHR\\ \hline
\end{tabular}
\caption{References of the components used for the setup to measure band structures.}
\label{tab:ref_BFP-Setup}
\end{table}

\medskip

To characterize the band structures of photonic crystals, reflectivity measurements were conducted under white light. It could also be characterized by transmissivity measurements, as the mirror downstream of the lens L\textsubscript{1} can be rotated. The position of the sample holder can be fine-tuned in all directions ($\vec{e}_x$, $\vec{e}_y$ and $\vec{e}_z$), and it can be rotated over 360°. The C\textsubscript{1} camera images the sample, allowing users to fine-tune the placement of photonic crystals. Once the latter is aligned, BS\textsubscript{2} can be flipped to increase the signal in the direction of the spectrometer. The slit filters the Fourier space in order to select a direction of high symmetry such as $\Gamma$X or $\Gamma$M for instance. Then the grating of the spectrometer disperses the filtered signal to image the band structure on the spectrometer detector.

\medskip

We introduce the azimuthal angle $\theta$ that refers to the light propagation angle in the direction of the slit, and we denote wavelengths by $\lambda$. The white light beam is considered uniform over the excited area of the sample, and its spectral irradiance is therefore denoted $I_{source}(\lambda)$. The transfer function of our setup taking into account the transmission of the different optics and the sensitivity of the detector, is denoted $S(\lambda, \theta)$. The reflectivity of a sample $i$, denoted $R_i(\lambda, \theta)$, is therefore linked to its measurement $M_i(\lambda, \theta)$ by the Equation \ref{eq:reflectivity_mes_1}, where $R_{background}(\lambda, \theta)$. Hence, the reflectivity of PhCs was deduced using a reference whose reflectivity $R_{ref}(\lambda, \theta)$ is known, and Equation \ref{eq:reflectivity_mes_2} where $t_i$ stand for the integration times of the PhC and the reference measurements.

\begin{equation}
    M_i(\lambda, \theta) = I_{source}(\lambda)S(\lambda, \theta)\times [R_i(\lambda, \theta) + R_{background}(\lambda, \theta)]
    \label{eq:reflectivity_mes_1}
\end{equation}

\begin{equation}
    R_{PhC}(\lambda, \theta) = \frac{M_{PhC}(\lambda, \theta) - M_{background}(\lambda, \theta)}{M_{ref}(\lambda, \theta) - M_{background}(\lambda, \theta)} \times \frac{t_{ref}}{t_{PhC}} \times R_{ref}(\lambda, \theta)
    \label{eq:reflectivity_mes_2}
\end{equation}

\subsection{Lattice parameter dependence of the band structure}
\label{app:lattice_parameter}

Based on the band structures of all photonic crystals measured experimentally by reflectivity, we observed the position of the TE\textsubscript{0}-like slow-light mode in the $\Gamma$-point depending on the lattice parameter (p). To determine the spectral evolution of the latter, we did a linear regression using the \textit{stats.linregress} function from \textit{scipy} package (1.16.0 version) in \textit{Python}. A correlation factor of r\textsuperscript{2} = 0.99 was found. We conclude that as the lattice parameter increases by 1 nm, the mode redshifts by 1.39 nm.

\begin{figure}[!h]
    \centering
    \includegraphics[]{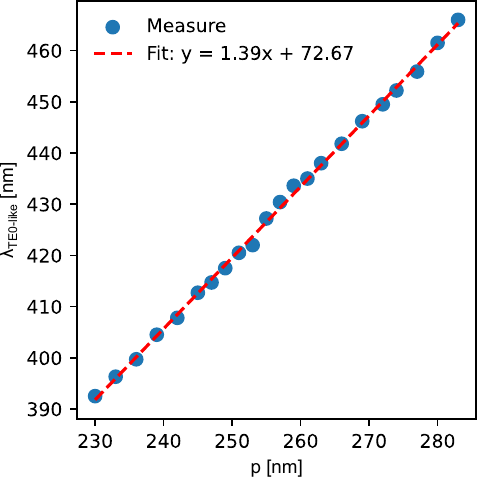}
    \caption{Experimental measurement of the TE\textsubscript{0}-like mode spectral position in the $\Gamma$-point depending on the lattice parameter of photonic crystals - Correlation factor r\textsuperscript{2} = 0.99}
    \label{fig:redshift}
\end{figure}
\FloatBarrier

\printbibliography[title={Supplemental References}]

\end{refsection}

\end{document}